\documentclass[prd,aps,singlecolumn,floatfix,preprintnumbers,nofootinbib]{revtex4}
\pdfoutput=1
\usepackage{subfigure}

\newcommand{\appropto}{\mathrel{\vcenter{\offinterlineskip\halign{\hfil$##$\cr
    \propto\cr\noalign{\kern2pt}\sim\cr\noalign{\kern-2pt}}}}}

\newcommand{\lsim}
{\;\raisebox{-.3em}{$\stackrel{\displaystyle <}{\sim}$}\;}

\usepackage{amsmath}
\usepackage{amssymb, bm}
\usepackage{graphicx}
\usepackage{float}
\usepackage{braket}
\usepackage{hyperref}
\usepackage{pbox}
\usepackage{tabularx}
\usepackage{enumitem}
\usepackage{tikz}\usetikzlibrary{intersections}
\usepackage{epstopdf}
\newcommand{\bmat}{\begin{pmatrix}}
\newcommand{\emat}{\end{pmatrix}}
\def\beq{\begin{equation}}
\def\eeq{\end{equation}}
\def\beeq{\begin{eqnarray}}
\def\eeeq{\end{eqnarray}}
\def\beqn{\begin{eqnarray}}
\def\eeqn{\end{eqnarray}}
\def\bea{\begin{align}}
\def\eea{\end{align}}

\def\gp2{g^{\prime 2}}

\usepackage{color}
\begin{document}   
\preprint{DAMTP-2015-92}
\preprint{IPPP/15/80}
\preprint{DCPT/15/160}
\title{750 GeV Di-photon Excess Explained by a Resonant Sneutrino in $R-$parity
  Violating Supersymmetry}  
\author{B.~C.~Allanach$^1$}
\author{P.~S.~Bhupal~Dev$^{2,3}$}
\author{S.~A.~Renner$^1$}
\author{Kazuki Sakurai$^4$} 
\affiliation{$^1$Department of Applied Mathematics and Theoretical Physics, Centre
  for Mathematical Sciences, University of Cambridge, Wilberforce Road,
  Cambridge CB3 0WA, UK}
\affiliation{$^2$Physik Department T30d,
Technische Universit\"{a}t M\"{u}nchen, James-Franck-Stra\ss e 1, D-85748
Garching, Germany}
\affiliation{$^3$Max-Planck-Institut f\"{u}r Kernphysik, Saupfercheckweg 1, D-69117 Heidelberg, Germany}
\affiliation{$^4$Institute for Particle Physics Phenomenology, Department of Physics, University of Durham, South Road, Durham DH1 3LE, UK}
\begin{abstract}
We explain the recent excess seen by ATLAS and CMS experiments at
around 750 GeV in the di-photon invariant mass as a narrow width sneutrino
decaying to di-photons
via a stau loop in $R-$parity violating Supersymmetry. The stau mass is
predicted to be somewhere between half the resonant sneutrino mass and half the sneutrino mass
plus 14 GeV. The scenario also predicts
further signal channels at an invariant mass of 750 GeV, the most promising
being into di-jets and $WW$. We also predict a left handed charged
slepton decaying into $WZ$ and $W \gamma$ at a mass 750-754 GeV.
\end{abstract}   
\maketitle 

\section{Introduction}

The ATLAS and CMS collaborations have recently presented the results of
di-photon resonance searches in early Run II of $\sqrt s=13$ TeV
data~\cite{ATLAS-CONF-2015-081, CMS:2015dxe, atlas13, CMS:2016owr}.  
For a spin-0 hypothesis, ATLAS observed an excess of 3.9 $\sigma$ local significance (2.0 $\sigma$
global) at a di-photon invariant mass of around 750 GeV with 3.2 fb$^{-1}$
integrated luminosity. 
CMS also observed a 2.9 $\sigma$ excess locally ($1.2$ $\sigma$ globally) at
a similar mass of 760 GeV in 3.3 fb$^{-1}$ of data. 
The ATLAS excess 
prefers a large width $\sim 45$ GeV, but only at a very mild level (the local
significance increases by 0.3\,$\sigma$ above the narrow width
approximation~\cite{atlas13}), whereas the CMS fit prefers a much narrower width~\cite{CMS:2016owr}. Together, these excesses are consistent with a new narrow-width resonance decaying into
two photons with a cross-section of 
$\sigma( pp \to \gamma \gamma) \approx 5.3 \pm 2.4$ fb 
(unfolding efficiency and acceptance as in
Ref.~\cite{Falkowski:2015swt}\footnote{This assumes efficiency times
  acceptance of 0.65 for ATLAS and 0.48 for CMS\@. These numbers were calculated
  assuming gluon fusion production, 
  which will not be our case. However, to the accuracy with which we work,
  the approximation should be sufficiently good.}).  
The possibility of a new 750 GeV resonance decaying into di-photons
has stimulated a lot of interesting ideas and speculations in the theory
community recently; for an incomplete list, see Refs.~\cite{
Harigaya:2015ezk,
Mambrini:2015wyu,
Backovic:2015fnp,
Angelescu:2015uiz,
Nakai:2015ptz,
Knapen:2015dap,
Buttazzo:2015txu,
Pilaftsis:2015ycr,
Franceschini:2015kwy,
DiChiara:2015vdm,
Higaki:2015jag,
Ellis:2015oso,
Low:2015qep,
Bellazzini:2015nxw,
Gupta:2015zzs,
Petersson:2015mkr,
Molinaro:2015cwg,
Dutta:2015wqh,
Cao:2015pto,
Matsuzaki:2015che,
Kobakhidze:2015ldh,
Martinez:2015kmn,
Cox:2015ckc,
Becirevic:2015fmu,
No:2015bsn,
Demidov:2015zqn,
Chao:2015ttq,
Fichet:2015vvy,
Curtin:2015jcv,
Bian:2015kjt,
Chakrabortty:2015hff,
Csaki:2015vek,
Ahmed:2015uqt,
Agrawal:2015dbf,
Falkowski:2015swt,
Aloni:2015mxa,
Bai:2015nbs,
Gabrielli:2015dhk,
Benbrik:2015fyz,
Kim:2015ron,
Alves:2015jgx,
Megias:2015ory,
Carpenter:2015ucu,
Bernon:2015abk,
Chao:2015nsm,
Arun:2015ubr,
Han:2015cty,
Chang:2015bzc,
Chakraborty:2015jvs,
Ding:2015rxx,
Han:2015dlp,
Han:2015qqj,
Luo:2015yio,
Chang:2015sdy,
Bardhan:2015hcr,
Feng:2015wil,
Antipin:2015kgh,
Wang:2015kuj,
Cao:2015twy,
Huang:2015evq,
Liao:2015tow,
Heckman:2015kqk,
Dhuria:2015ufo,
Bi:2015uqd,
Kim:2015ksf,
Berthier:2015vbb,
Cho:2015nxy,
Cline:2015msi,
Bauer:2015boy,
Chala:2015cev,
Barducci:2015gtd,
Boucenna:2015pav, 
Murphy:2015kag, 
Hernandez:2015ywg, 
Dey:2015bur, 
Pelaggi:2015knk, 
deBlas:2015hlv, 
Dev:2015isx,
Huang:2015rkj, 
Moretti:2015pbj, 
Patel:2015ulo, 
Badziak:2015zez, 
Chakraborty:2015gyj,
Cao:2015xjz, 
Altmannshofer:2015xfo, 
Cvetic:2015vit, 
Gu:2015lxj}. Many of the interpretations rely on heavy Higgs or other scalar
bosons with additional charged particles that enhance the di-photon branching
ratio and the total width. 

\begin{figure}[t!]
  \centering 
    \includegraphics[width=0.35\textwidth]{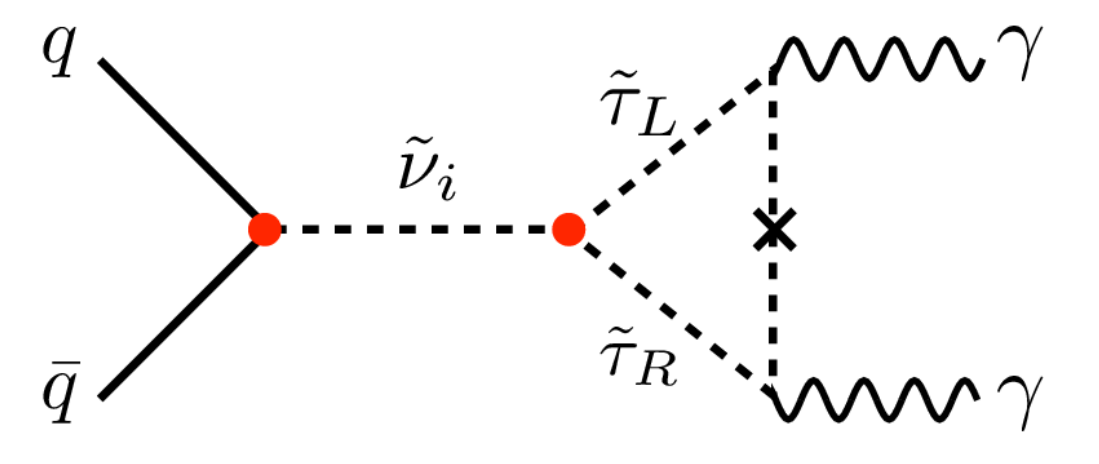}
\hspace{0.5cm}
\includegraphics[width=0.35\textwidth]{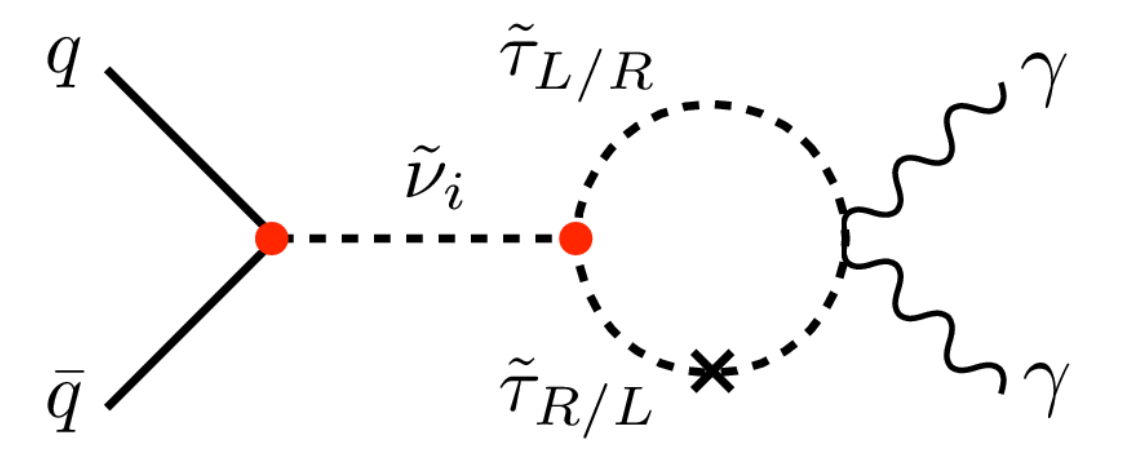}
    \caption{Example Feynman diagrams for resonant sneutrino production via the
      $L_i  Q_1 \bar{D}_1$ 
      operator in Eq.~\eqref{eq:wrpv} and its decay to two photons via the soft term $\tilde \ell_i \tilde \ell_3
      \tilde \tau_R^+$ in Eq.~\eqref{eq:lsoft}. There are two kinds of diagrams: {\it (left)} through the
      triangle stau loop, and {\it (right)} through the $\tilde
\tau_{R/L} \tilde \tau_{L/R}^* \gamma \gamma$ vertex, which must be included in the calculation to cancel the divergences in the loop integrals. 
    The cross in the stau propagators represents the left-right mixing in the
    stau sector, which must be non-zero to have a di-photon signal.
\label{fig:main_diag}}
\end{figure}

In this work we interpret the observed di-photon excess {\it within} the Minimal
Supersymmetric Standard Model (MSSM) framework as a 750
GeV scalar neutrino (sneutrino) resonance, 
$d \bar d \to \tilde \nu_i$, produced via the R-parity violating (RPV)
interaction
\beq
W_{\rm LV} \  =  \ \lambda'_{i11} L_i Q_1 \bar D_1
\; ,  \\ 
\label{eq:wrpv}
\eeq
where $i$ is the family index of the sneutrino.
The sneutrino may decay into two photons through a stau loop with a left-right stau mixing via the RPV soft supersymmetry (SUSY) breaking term
\beq
{\cal L}^{\rm soft}_{\rm LV} \ = \ A_{i33} \tilde \ell_i \tilde \ell_3 \tilde
\tau_R^+ + {\rm (H.c.)}\; , \label{eq:lsoft}
\eeq
where the $SU(2)_L$ indices of $\tilde \ell_i$ and $\tilde \ell_3$ are
anti-symmetrically contracted implicitly, which forbids $i$ to be 3, so 
the 750 GeV sneutrino has to be of electron or muon type in our scenario.
There are two kinds of stau loops, as shown in Fig.~\ref{fig:main_diag}, that will contribute to the di-photon signal and may explain the excesses observed in the
ATLAS and CMS data, as shown below. Assuming that 
the resonance is a heavy neutral Higgs boson of the
 MSSM, the
 production cross section prediction is too small~\cite{Angelescu:2015uiz}
   unless additional non-MSSM states are added.\footnote{However, neutral
     Higgs bosons in the NMSSM could explain the di-photon
     excess~\cite{Ellwanger:2016qax, Domingo:2016unq,
       Badziak:2016cfd}. Another interesting possibility in spontaneously
     broken SUSY models is the sgoldstino~\cite{Bellazzini:2015nxw,
       Petersson:2015mkr, Demidov:2015zqn, Casas:2015blx, Ding:2016udc,
       Bardhan:2016rsb}.} Thus, our interpretation in terms of one of the {\it
     only} other viable neutral 
scalars in the MSSM, namely a sneutrino, should serve as a well-motivated
and minimal solution.

The rest of the paper is organised as follows.
In Sec.~\ref{sec:decay} we consider the decay of the sneutrino and discuss the constraints on our scenario.
In Sec.~\ref{sec:result} we show our results and discuss the value of the sneutrino width that one can obtain 
in our scenario. In Sec.~\ref{sec:low} we show that all the relevant low-energy constraints can be satisfied for the di-photon favoured region in this model.  Sec.~\ref{sec:tweaks} discusses how one might tweak the
model in order to increase the width of the sneutrino in the event that it is
unambiguously measured by the experiments to be a wide resonance. Sec.~\ref{sec:concl} is devoted to
conclusions.

\section{Sneutrino decay} \label{sec:decay}

Given the interaction terms in Eq.~\eqref{eq:wrpv}, 
the sneutrino $\tilde \nu_i$ of mass  $m_{{\tilde \nu}_i}$
may decay into $d \bar d$ with the following
partial width: 
\beq
\Gamma_{d \bar d} \ \equiv \ \Gamma(\tilde \nu_i \to d \bar d) \ = \ \frac{3 }{16 \pi} |\lambda_{i11}^\prime|^2 m_{\tilde \nu_i}.
\label{eq:dec_dd}
\eeq  
This decay is unavoidable because it is the inverse process to the production,
and is constrained  
by the di-jet resonance searches \cite{CMS:2015neg, Aad:2014aqa}. 
With the interaction terms in Eq.~\eqref{eq:lsoft}, 
the sneutrino may also decay into a pair of staus if $2 m_{\tilde \tau_1} \leq m_{\tilde \nu_i}$,
where $\tilde \tau_1$ is the lighter mass-eigenstate of the staus, with
partial width
\beeq
\Gamma_{\tilde \tau \tilde \tau} \ \equiv \ \Gamma( \tilde \nu_i \to \tilde \tau_1^+ \tilde \tau_1^- ) \  = \  \frac{|A_{i33}|^2}{16 \pi m_{\tilde \nu_i}} \left(1-\frac{4m^2_{\tilde{\tau}_1}}{m_{\tilde \nu_i}^2}\right)^{1/2},
\label{eq:dec_stau}
\eeeq 
where $m_{\tilde \tau_1}$ is the mass of the lightest stau.
 In this case, the branching ratio to the loop-induced di-photon decay mode shown in Fig.~\ref{fig:main_diag} will be hugely suppressed, thus disfavouring the di-photon signal. 
However, if the decay to on-shell staus is kinematically impossible ($2 m_{\tilde
  \tau_1} > m_{\tilde \nu_i}$) 
and the hadronic decay in Eq.~\eqref{eq:dec_dd} is suppressed
($|\lambda_{i11}^\prime| \ll 1$),  
the sneutrino can decay with an appreciable branching ratio into  neutral
gauge bosons $\gamma \gamma, \gamma 
Z, ZZ$ 
via the one-loop diagram of the staus shown in Fig.~\ref{fig:main_diag}. 
Neglecting the contribution from the heavier state $\tilde \tau_2$,
 the partial 
widths are given by 
\beeq
\Gamma_{\gamma \gamma} \ \equiv \ \Gamma(\tilde \nu_i \to \gamma \gamma) & \ = \ & \frac{\alpha^2 m^3_{\tilde \nu_i} }{256 \pi^3} 
\frac{ |\bar A_{i33}|^2}{ m^4_{\tilde \tau_1} }  
\left|A_0( \tau_{\tilde \tau} )\right|^2, \label{eq:dec_gg}
\\
\Gamma_{\gamma Z} \ \equiv \ \Gamma(\tilde \nu_i \to \gamma Z) & \ = \ & \frac{\alpha^2 m^3_{\tilde \nu_i} }{128 \pi^3} 
\frac{ |\bar A_{i33}|^2 }{ m^4_{\tilde \tau_1} } \left( 1 - \frac{m^2_Z}{m^2_{\tilde \nu_i}} \right)^3 
\left|\lambda_{Z\tilde \tau_1\tilde \tau_1} A_{0Z}(\tau^{-1}_{\tilde \tau}, \tau^{-1}_Z )\right|^2,
\\
\Gamma_{ZZ} \ \equiv \ \Gamma(\tilde \nu_i \to Z Z) & \ = \ & \frac{\alpha^2 m^3_{\tilde \nu_i} }{256 \pi^3} \frac{ |\bar A_{i33}|^2}{ m^4_{\tilde \tau_1} }  \left( 1 - \frac{4 m^2_Z}{m^2_{\tilde \nu_i}} \right)^3 
\left|\lambda^2_{Z\tilde \tau_1\tilde \tau_1} A_{0Z}( \tau^{-1}_{\tilde \tau}, \tau^{-1}_Z )\right|^2, 
\label{eq:dec_ZZ}
\eeeq
where 
$\bar A_{i33} \equiv A_{i33} \cos \theta \sin \theta$,
$\lambda_{Z\tilde \tau_1\tilde \tau_1} \equiv a-b \cos 2\theta$, 
with $a \equiv (3\tan\theta_w-\cot\theta_w)/4$, $b\equiv (\tan\theta_w+\cot\theta_w)/4$, $\theta_w$ being the weak mixing angle and 
$\theta$ being the left-right mixing angle of the stau sector:
i.e.~$\tilde \tau_{R/L} = \tilde \tau_{1/2} \cos \theta  \pm  \tilde \tau_{2/1} \sin \theta$.  
Also, $\tau_{\tilde \tau} \equiv m^2_{\tilde \nu_i}/4 m^2_{\tilde \tau_1}$,
$\tau_Z \equiv m^2_Z/4 m^2_{\tilde \tau}$ and
the scalar loop functions $A_0$ and $A_{0Z}$ are defined by
\beqn
A_0(x)& \ = \ &-\frac{x-f(x)}{x^2} \, , \\ 
A_{0Z}(x_1, x_2) &  = &\frac{x_1 x_2}{2(x_1-x_2)} 
+\frac{x_1^2 x_2^2}{2(x_1 - x_2)^2}\,\left[f(x_1^{-1}) - f(x_2^{-1})\right] \nonumber \\ 
& & 
+  \frac{x_1^2 x_2}{(x_1 - x_2)^2}\, \left[(g(x_1^{-1}) - g(x_2^{-1})\right], 
\label{eq:loop_func}
\eeqn
where the functions $f$ and $g$ are
\begin{align}
f(x) \ & = \ 
\left\{
    \begin{array}{ll}
        \arcsin^2(\sqrt{x})  & \mbox{if } x \leq 1 \\
        -\frac{1}{4}\left[\log\left(\frac{1 + \sqrt{1 - 1/x}}{1 - \sqrt{1 - 1/x}}\right) - 
    i\pi\right]^2& \mbox{if } x > 1  \, ,
    \end{array}
\right. \\ 
g(x) \ & =
\ \left\{
    \begin{array}{ll}
                \frac{\sqrt{1 - 1/x}}{2}\left[\log\left(\frac{1 + \sqrt{1 - 1/x}}{1 - \sqrt{1 - 1/x}}\right) - i\pi\right]& \mbox{if } x < 1 \\
                \sqrt{1/x-1}\: \arcsin(\sqrt{x})  & \mbox{if } x \geq 1 \, .
    \end{array}
\right. 
\end{align}
One can see that these partial widths are proportional to $\sin 2 \theta$
through $\bar A_{i33}$, 
meaning that a large left-right mixing is required to obtain
a large di-photon branching ratio.
This can also be understood diagramatically due to the presence of  the cross on the stau propagator in Fig.~\ref{fig:main_diag}.\footnote{In principle, one can also allow for a large mixing in the selectron or smuon sector with a large $\mu$-term and/or large $\tan\beta$, albeit with some tuning of the parameters to avoid tachyonic states. Our subsequent analysis is equally applicable to these cases, but we stick to staus for definiteness.} 
If the stau sector has a large left-right mixing, one tends to have
a large mass hierarchy, $m_{\tilde \tau_2} \gg
m_{\tilde \tau_1}$. We can therefore neglect the $\tilde \tau_2$
  contribution in the loop. 
On the other hand, the $\tilde \nu_\tau$
  contribution relevant for the $\tilde{\nu}_i\to W^+W^-$ decay mode
  through the $\tilde\tau_L-\tilde\tau_R-\tilde\nu_\tau$ triangle loop need not be
  negligible in the large mixing limit. To be precise, the $WW$ partial width
  in the limit $m_{\tilde \tau_2} \gg m_{\tilde
    \tau_1}$ is given by  
\begin{align}
& \Gamma_{WW} \  \equiv  \ \Gamma(\tilde \nu_i \to W^+W^-)  \ = \  \frac{\alpha_w^2 m^3_{\tilde \nu_i} }{1024 \pi^3} \frac{ |\bar A_{i33}|^2}{ m^4_{\tilde \tau_1} } \sin^4\theta \left( 1 - \frac{4 m^2_W}{m^2_{\tilde \nu_i}} \right)^{1/2} 
  \nonumber \\ 
&\times 
\bigg[ \frac{|F|^2}{16\tau^2_{\tilde \tau}}\left(12-\frac{4m^2_{\tilde \nu_i}}{m_W^2}+\frac{m^4_{\tilde \nu_i}}{m_W^4} \right) 
-\frac{|F\cdot G| }{2\tau_{\tilde \tau}}\left(8-\frac{6m^2_{\tilde \nu_i}}{m_W^2}+\frac{m^4_{\tilde \nu_i}}{m_W^4}   \right)+|G|^2\left(16-\frac{8m^2_{\tilde \nu_i}}{m_W^2}+\frac{m^4_{\tilde \nu_i}}{m_W^4} \right) \bigg], 
\label{eq:dec_WW}
\end{align}
where $\alpha_w\equiv g_w^2/4\pi$, $g_w$ is the $SU(2)_L$ gauge coupling, and 
\begin{align}
F(m^2_{\tilde \nu_i}, m^2_{\tilde \tau_1},m^2_{\tilde \nu_\tau},m^2_W)  \ = \ & 2C_{00}(m^2_{\tilde \nu_i}, m^2_W, m^2_W, m^2_{\tilde \tau_1},m^2_{\tilde \tau_1},m^2_{\tilde \nu_\tau})-\frac{1}{2}B_0(m^2_{\tilde \nu_i}, m^2_{\tilde \tau_1},m^2_{\tilde \tau_1}), \\
G(m^2_{\tilde \nu_i}, m^2_{\tilde \tau_1},m^2_{\tilde \nu_\tau},m^2_W)  \ = \ &  
m^2_{\tilde \tau_1}\big[C_{11}(m^2_{\tilde \nu_i}, m^2_W, m^2_W, m^2_{\tilde \tau_1},m^2_{\tilde \tau_1},m^2_{\tilde \nu_\tau}) \nonumber \\
& \qquad +C_{12}(m^2_{\tilde \nu_i}, m^2_W, m^2_W, m^2_{\tilde \tau_1},m^2_{\tilde \tau_1},m^2_{\tilde \nu_\tau}) \nonumber \\
& \qquad +C_{1}(m^2_{\tilde \nu_i}, m^2_W, m^2_W, m^2_{\tilde \tau_1},m^2_{\tilde \tau_1},m^2_{\tilde \nu_\tau}) \big]
\label{eq:Floop}
\end{align}
with $B_0$, $C_1$, $C_{00,11,12}$ being the usual scalar two- and three-point Passarino-Veltman functions~\cite{Passarino:1978jh} in the conventions of Ref.~\cite{Denner:1991kt}, which we evaluate numerically using {\tt
  LoopTools}~\cite{Hahn:1998yk}. 
From Eqs.~\eqref{eq:dec_gg} and~\eqref{eq:dec_WW}, we find that the $WW$ partial width can be suppressed with respect to the di-photon width by a suitable choice of the mass and mixing parameters in the stau sector.

The total decay width $\Gamma_{\rm tot}$ of the sneutrino in our scenario is thus 
given by 
\beq
\Gamma_{\rm tot} \ \simeq \ \Gamma_{d \bar d}+ \Gamma_{\tilde \tau \tilde\tau}+ \Gamma_ {\gamma \gamma} +  \Gamma_{\gamma Z} + \Gamma_{ZZ} + \Gamma_{WW}+ \Gamma_{X},
\eeq
where the partial widths are given in Eqs.~\eqref{eq:dec_dd}-\eqref{eq:dec_ZZ} and \eqref{eq:dec_WW}, and  $\Gamma_{X}$ is
the contribution from any other possible decay channels not explicitly
mentioned  here but that could potentially have an appreciable partial width
(by changing model parameters and making other super partners non-decoupled).

\begin{figure}[t!]
  \centering 
\includegraphics[width=0.45\textwidth]{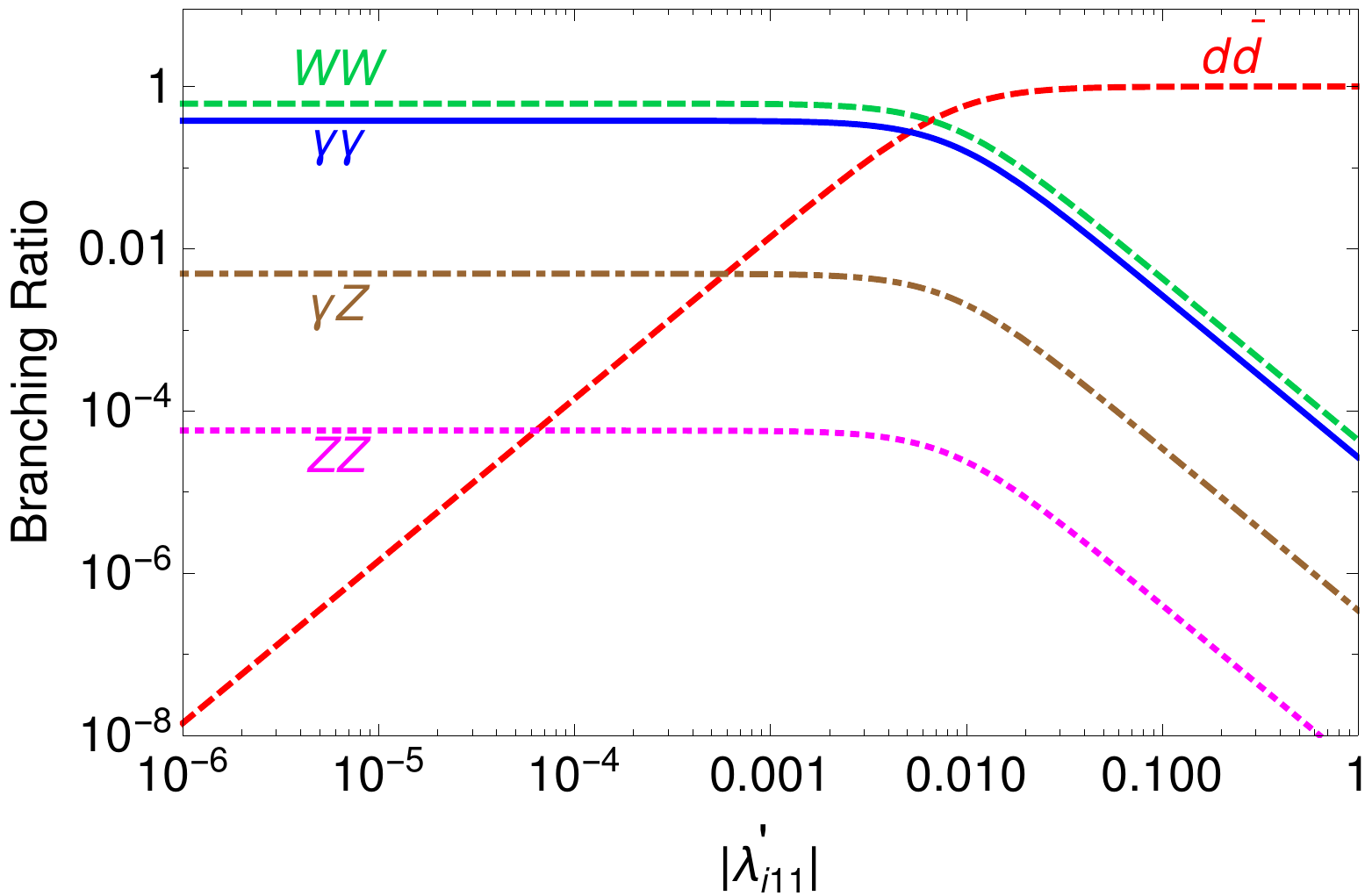}
\caption{The branching ratios of the sneutrino decay 
      to  $d \bar d$, $\gamma \gamma$, $\gamma Z$, $ZZ$ and $WW$. Here we have
    chosen the benchmark values given in Eq.~\eqref{bench2}, in addition to setting $m_{\tilde \nu_i} = 750$ GeV and $A_{i33}=14\: m_{\tilde\tau_1}$. 
\label{fig:br}}
\end{figure}

For a numerical illustration, we choose the following benchmark values for the stau sector: 
\begin{align}
\tilde m^2_{L_3} \ & = \ m_{\tilde \ell_3} + m^2_{\tau} + m_Z^2 \cos 2 \beta (-\frac{1}{2} + \sin^2 \theta_w ) \ = \ (425~{\rm GeV})^2, \nonumber \\  
\tilde m^2_{R_3} \ & = \ m^2_{\tilde \tau_R} + m^2_{\tau} - m_Z^2 \cos 2 \beta \sin^2 \theta_w \ = \ (445~{\rm GeV})^2, \nonumber \\
X_\tau \ & = \ m_\tau(A_\tau-\mu\tan\beta) \ = \ -43~{\rm GeV}^2, \nonumber \\
\tan\beta \ & = \ 20.
\label{bench1}
\end{align}
The stau mass-squared matrix in the gauge eigenbasis $(\tilde\tau_L,\tilde\tau_R)$ is given by 
\begin{align}
M^2_{\tilde\tau} \ = \ \left( \begin{array}{cc}
\tilde m_{L_3}^2 & X_\tau \\
X_\tau & \tilde m_{R_3}^2
\end{array}\right),
\label{staumass}
\end{align}
with the left-right stau mixing given by $\tan 2\theta = 2X_\tau/(\tilde m^2_{L_3}-\tilde m^2_{R_3})$. The tau sneutrino mass is given by $m^2_{\tilde\nu_\tau} = m^2_{\tilde \ell_3}+(1/2)m_Z^2\cos 2\beta$. Thus, Eqs.~\eqref{bench1} lead to the following mass and mixing values: 
\begin{align}
m_{\tilde\tau_1} \ = \ 382~{\rm GeV},\quad m_{\tilde\tau_2} \ = \ 483~{\rm GeV}, \quad m_{\tilde\nu_\tau} \ = \ 416~{\rm GeV}, \quad \sin^2\theta \ = \ 0.4.
\label{bench2}
\end{align}
We now compute the branching ratios of the
sneutrino decay to di-jet, di-photon, $\gamma Z$, $ZZ$ and $WW$ channels using Eq.~\eqref{bench2}. This is
shown in Fig.~\ref{fig:br} for a suitable choice of parameters $m_{\tilde
  \nu_i} = 750$ GeV and $A_{i33}=14\: m_{\tilde\tau_1}$.  
From Fig.~\ref{fig:br}, we
find that the di-photon branching ratio is sizable for small
$\lambda'_{i11}$, which however cannot be made arbitrarily small, since the
sneutrino production cross section is proportional to $|\lambda'_{i11}|^2$. We
also note that the partial widths for  
$\tilde \nu_i \to \gamma Z$ and $\tilde \nu_i \to Z Z$
are respectively $\sim 10^{-2}$ and $\sim 10^{-4}$ of 
$\Gamma_{\gamma \gamma}$. 

\begin{figure}[t!]
  \centering 
\includegraphics[width=0.45\textwidth]{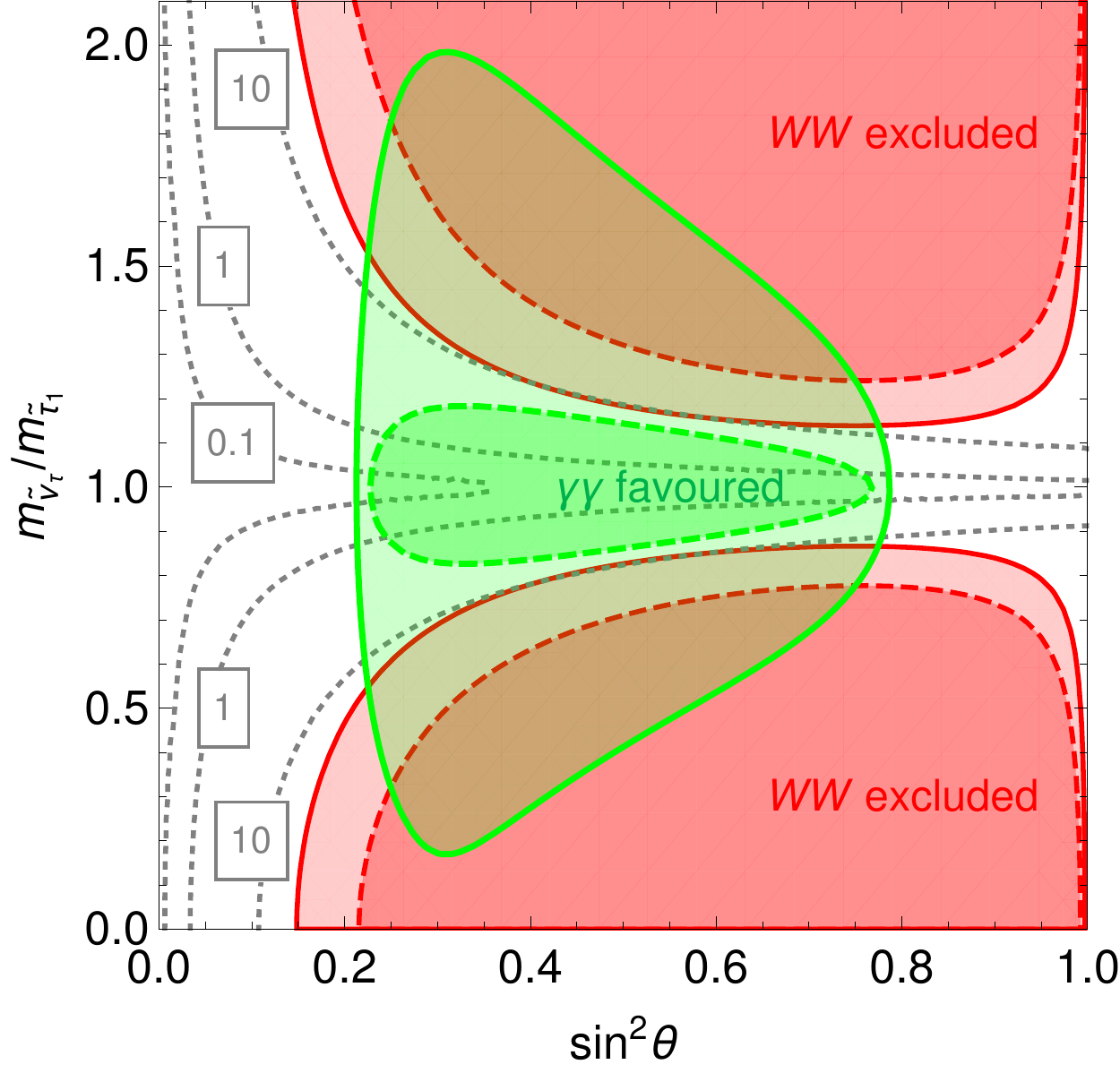}
\caption{The contours of ${\rm BR}_{WW}/{\rm BR}_{\gamma \gamma}$ (dotted curves) as a function of the stau mixing angle $\theta$ and the mass 
      ratio $m_{{\tilde \nu}_\tau}/{m_{{\tilde \tau}_1}}$ for fixed values of 
      $m_{{\tilde \tau}_1}=380$ GeV, $m_{\tilde \nu_i} = 750$ GeV and
      $A_{i33}=14\: m_{\tilde\tau_1}$. The red-shaded regions enclosed by the
      red solid (dashed) curves are the 95\% CL exclusion regions for
      $|\lambda'_{i11}|=0.08~(0.02)$ from the $\sqrt s=8$ TeV LHC $WW$
      data. The green-shaded regions enclosed by the green solid (dashed)
      curves are the $1\sigma$ favoured regions for
      $|\lambda'_{i11}|=0.08~(0.02)$ to explain the $\sqrt s=13$ TeV LHC
      di-photon excess.} 
\label{fig:WW}
\end{figure}

On the other hand, the $WW$ partial width can
be comparable to or larger than the di-photon width, depending on the stau
mixing and tau 
sneutrino mass, as depicted in Fig.~\ref{fig:WW}. In particular, for smaller
stau mixing, the $WW$ rate is 
suppressed with respect to the $\gamma\gamma$ due to the additional  
$\sin^4\theta$ dependence in Eq.~\eqref{eq:dec_WW}, but we cannot take the
mixing to be arbitrarily small, as it would also suppress the $\gamma\gamma$
rate with respect to the di-jet rate. We find that $\theta$ must be between $\pi/7$ and $\pi/3$ to have a di-photon favoured region consistent with other
constraints (see Section~\ref{sec:result}). Similarly, if the tau sneutrino mass
is close to the stau mass, the ratio ${\rm BR}_{WW}/{\rm BR}_{\gamma
  \gamma}$ is small, giving a larger parameter space for the di-photon
signal. In Fig.~\ref{fig:WW}, the red-shaded regions enclosed by the red solid
(dashed) curves are the 95\% CL exclusion regions for
$|\lambda'_{i11}|=0.08~(0.02)$ from the $\sqrt s=8$ TeV LHC $WW$
data~\cite{Khachatryan:2015cwa, Aad:2015agg}. The green-shaded regions
enclosed by the green solid (dashed) curves are the $1\sigma$ favoured regions
for $|\lambda'_{i11}|=0.08~(0.02)$ to explain the $\sqrt s=13$ TeV LHC
di-photon excess~\cite{atlas13, CMS:2016owr}. Here, we cannot take a larger
value of $|\lambda'_{i11}|$, otherwise it will be in conflict with the $\sqrt
s=8$ TeV di-jet constraints~\cite{CMS:2015neg, Aad:2014aqa} (see
Fig.~\ref{fig:RPV0}). A smaller value of $|\lambda'_{i11}|$ will give a
smaller di-photon favoured region. We see from the figure that there is ample
room in parameter space where a resonant sneutrino fits the di-photon excess
whilst simultaneously satisfying constraints on resonant $WW$ production.    

\section{Results} \label{sec:result}

\begin{figure}[t!]
  \centering 
    \includegraphics[width=0.4\textwidth]{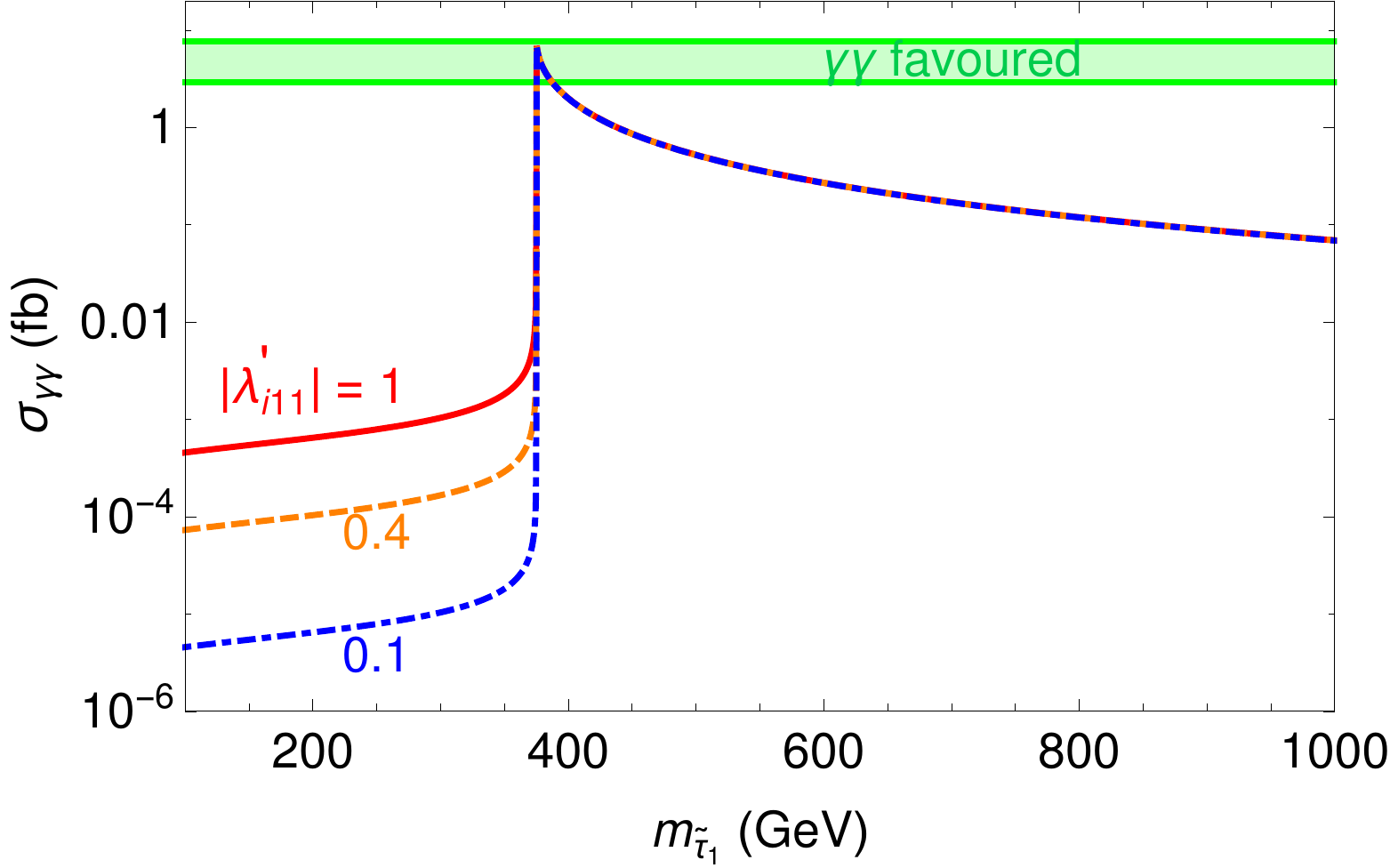}
    \caption{The di-photon signal cross section times branching ratio as a
      function of the stau mass for different values of the RPV coupling
      $\lambda'_{i11}$. \label{fig:sig}}  
\end{figure}
We compute the signal cross section at $\sqrt s=13$ TeV LHC using the RPV
model implementation in {\tt FeynRules}~\cite{Alloul:2013bka} and the
parton-level event generation in {\tt MadGraph5}~\cite{Alwall:2014hca} with
{\tt NNPDF2.3} leading order PDF sets~\cite{Ball:2012cx}. We find 
\beqn
\sigma(pp \to \tilde \nu_i \to \gamma \gamma)_{13 {\rm TeV}} 
& \ = \ & 
\sigma_0^{13 {\rm TeV}} |\lambda^\prime_{i11}|^2 \cdot {\rm BR}_{\gamma \gamma} \,,
\eeqn
where $\sigma_0^{13 {\rm TeV}} = 156$ pb for $m_{\tilde \nu_i} = 750$ GeV with $\lambda^\prime_{i11} = 1$. 
We require that the signal cross section be within the $1\sigma$ region of
the  observed value, i.e. $5.3 \pm 2.4$ fb~\cite{Falkowski:2015swt}.
Fig.~\ref{fig:sig} shows predictions for the signal cross section times
branching 
ratio as a function of the lightest stau mass for different values of
$\lambda'_{i11}$. When the stau mass is smaller than half the resonant
sneutrino mass at the left-hand side of the plot, the branching ratio to the
di-photon channel is highly suppressed and consequently the signal
cross section is much too small. 
It is clear from the figure that when the stau mass is
half (or just over half) the resonant sneutrino mass, the cross section fits
the di-photon excess measurements. Here, on-shell stau production is
kinematically disfavoured, boosting the $\gamma \gamma$ branching ratio, but
as the stau mass further increases, the loop diagram depicted in
Fig.~\ref{fig:main_diag} becomes increasingly mass suppressed and the signal
cross section dies off. 

There exist constraints on the di-boson decay modes from the 8 TeV LHC
data~\cite{Franceschini:2015kwy}. For the benchmark point shown in
Figure~\ref{fig:br}a, all these constraints are satisfied, except that there
is a small $2\sigma$ level tension in the $\gamma\gamma$ channel between the
Run-I and Run-II data sets for the production mode through $d\bar{d}$
annihilation, as considered here. 

On the other hand, the $\tilde \nu_i \to d \bar d$ channel is constrained by
the di-jet resonance searches \cite{CMS:2015neg, Aad:2014aqa}.  
The most stringent constraint comes from the $\sqrt s=8$ TeV LHC
data~\cite{Khachatryan:2016ecr}\footnote{Note that the reported results from
  the early Run II LHC di-jet 
resonance searches~\cite{ATLAS:2015nsi, Khachatryan:2015dcf} do not cover the
region at di-jet invariant masses of 750 GeV at all.}:
\beqn
\sigma(pp \to \tilde \nu_i \to d \bar d)_{8 {\rm TeV}} 
\  \simeq  \ \sigma_0^{8 {\rm TeV}} |\lambda^\prime_{i11}|^2 \cdot {\rm BR}_{d \bar d}  \ \lsim \ 0.9 \, {\rm pb}, 
\eeqn
where $\sigma_0^{8 {\rm TeV}} = 57$ pb is the $\sqrt s=8$ TeV production cross
section  for $pp \to \tilde \nu_i$ with $\lambda_{i11}^\prime = 1$ and
${\rm BR}_{d \bar d} = \Gamma_{d \bar d}/\Gamma_{\rm tot}$ is the branching ratio of the di-jet decay mode.  

Let us first consider the $\Gamma_{X} = 0$ case.
Since the upper limit of the di-jet cross section (0.9 pb) is much larger than the preferred di-photon cross section (8 fb),
we have $\Gamma_{d \bar d} \gg \Gamma_{\gamma \gamma}$ in the most of the interesting parameter region.
In this regime the total width of the sneutrino can be approximated by $\Gamma_{d \bar d} \appropto |\lambda_{i11}'|^2$ and we have
\beqn
\sigma(pp \to \tilde \nu_i \to \gamma \gamma) \ &\appropto& \ |\lambda^\prime_{i11}|^2 \cdot \left( \frac{\Gamma_{\gamma \gamma} }{|\lambda^\prime_{i11}|^2} \right) 
 \propto  \Gamma_{\gamma \gamma}, \\
\sigma(pp \to \tilde \nu_i \to d \bar d) \ &\appropto& \ |\lambda^\prime_{i11}|^2 \cdot \left( \frac{\Gamma_{d \bar d} }{|\lambda^\prime_{i11}|^2} \right) 
 \propto  |\lambda^\prime_{i11}|^2.
 \label{eq:snu}
\eeqn
Thus, the $\gamma \gamma$ signal rate is approximately independent of
$\lambda_{i11}'$ as 
Fig.~\ref{fig:sig} shows in the region $m_{{\tilde \tau}_1} > m_{{\tilde
    \nu}_i}/2$.

\begin{figure}[t!]
  \centering 
    \includegraphics[width=0.4\textwidth]{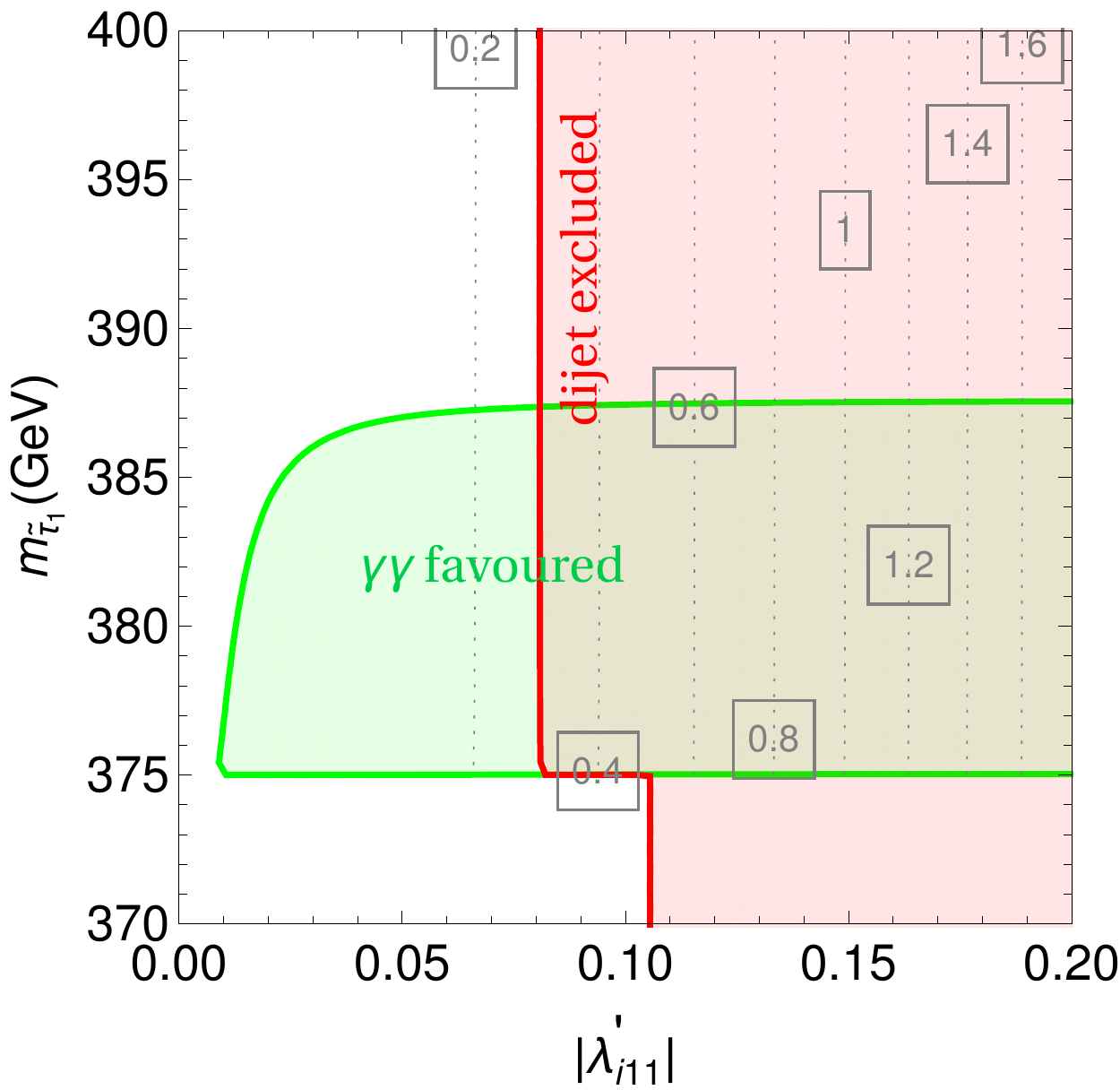}
    \caption{The preferred region for the di-photon excess (green) and
    the excluded region from the di-jet resonance search (red),
    assuming there is no other decay channel than 
    $d \bar d$, $\gamma \gamma$, $\gamma Z$, $ZZ$ and $WW$.
    The dashed contours show the total decay width of the sneutrino in GeV.
\label{fig:RPV0}}
\end{figure}

The di-jet signal cross section also receives a contribution from charged
slepton 
production:
\beqn
\sigma(pp \rightarrow {\tilde e}^-_{L_i} \rightarrow \bar u d) & \ = \ &
\sigma_-^\textrm{8~TeV} | \lambda'_{i11} |^2 
{\rm BR}(\tilde e_{L_i}^- \to \bar u d),
\label{eq:e-}
\\
\sigma(pp \rightarrow {\tilde e}^+_{L_i} \rightarrow u \bar d) & \ = \ &
\sigma_+^\textrm{8~TeV} | \lambda'_{i11} |^2
{\rm BR}(\tilde e_{L_i}^+ \to u \bar d).
\label{eq:e+}
\eeqn
In the following discussion, we assume ${\rm BR}(\tilde e_{L_i}^{\pm} \to u \bar d/\bar u d) = 1$ for simplicity.
This leads to an conservative upper limit on $|\lambda'_{i11}|$ from the
  di-jet constraint, which could be somewhat relaxed for ${\rm BR}(\tilde
e_{L_i}^{\pm} \to u \bar d/\bar u d) < 1$. 
We also assume that the LR mixing in the $\tilde e^{\pm}_i$ sector is negligible.
This is justified since the LR mixing is proportional to the fermion mass, which is negligible for first two generations.
At tree level, we have
$m^2_{{\tilde e}_{L_i}} =
m_{{\tilde \nu}_i}^2 - M_W^2 \cos 2 \beta$; 
thus in the range of $\beta \in [ \pi/4,\ \pi/2]$ considered,
$750<m_{{{\tilde e}_i}}/\text{~GeV}< 754$
for $m_{{\tilde \nu}_i}=750$ GeV.
This means that the sum of Eqs.~\eqref{eq:snu}, \eqref{eq:e-} and \eqref{eq:e+}
is constrained by the di-jet bound: we have included each in the calculation
of the bound in Fig.~\ref{fig:RPV0}.
We obtain $\sigma_-^\textrm{8~TeV}=23$ pb and $\sigma_+^\textrm{8~TeV}=57$
pb for a 750 GeV charged slepton. The charged slepton has decays into
$W\gamma$ or $WZ$ via a loop of stau/stau-sneutrino, with expected partial
widths of the same order as the $WW$ and $\gamma \gamma$ partial width of the
sneutrino. These channels therefore bring additional verification
possibilities. 


Fig.~\ref{fig:RPV0} shows our numerical result for the $\Gamma_X = 0$ case. 
Throughout this section, we take the near-maximal left-right mixing with $\sin^2\theta = 0.4$
and $A_{i33} = 14 \: m_{\tilde \tau_1}$ so that the signal rate is enhanced without too much fine-tuning.\footnote{The signal rate is maximized for $\theta=\pi/4$ and $m_{\tilde\nu_\tau}=m_{\tilde\tau_1}$ which is possible, but requires a large fine-tuning with $\tilde m^2_{L_3}=\tilde m^2_{R_3}$ in Eq.~\eqref{staumass}. 
Also note that the chosen value of $A_{i33}$ is roughly at the upper limit from perturbativity arguments~\cite{Allanach:2015blv}: larger values of $A_{i33}$ generate a
 large $|\tilde \ell_i|^4$ operator 
via a one-loop box diagram involving a loop of $\tilde \tau_1$
\cite{Allanach:2015blv}. 
There have been arguments proposed (see e.g.~\cite{Salvio:2016hnf}) to the
effect that such
a large 
trilinear coupling may destabilise the potential at large field values of
$\tilde \tau_1$. In this case, additional (non-MSSM) heavy states would be
required to modify the high energy behaviour of the potential such
that the stability of the vacuum is restored, even with the large value of $A_{i33}$
chosen here. 
Another resolution of the stability would be to lower $A_{i33}$ and enhance
the signal rate by using additional sneutrino states in the loop. 
For instance, if $m_{\tilde \nu_e} \simeq m_{\tilde \nu_\mu} \simeq  750$ GeV, 
one can obtain an enhancement factor of $2^2 = 4$, although in this case the mass splitting between the sneutrinos would need to be smaller than their
${\cal O}$(1 GeV) widths.
One could also have additional enhancement from a smuon contribution in the
loop, where non-zero $A_{i22}$ as well as a light ${\tilde \mu_1}$ with a large
mixing between left and right handed smuons would be required.
Yet another possibility is to produce $\tilde \nu_i$ from a cascade decay of $e^{\pm}_{L_i}$ as 
$pp \to \tilde e^{\pm}_{L_i} \to W^{\pm} \tilde \nu_i$.
This requires large LR mixing in the light-flavour slepton sectors
in order to have a large enough mass splitting between $\tilde e^{\pm}_{i}$
and $\tilde \nu_i$ 
to allow this decay.
Since the current di-photon excess has only $\sim 20$ events,
having this sub-leading contribution with an extra $W$ boson is still consistent with data.
In this case, the di-jet constraint is also relaxed because 
${\rm BR}(\tilde e_{L_i}^{\pm} \to u \bar d/\bar u d) < 1$.
}
In the green shaded region, the di-photon signal rate is within the 1 $\sigma$
band of the observed value, 
whereas the red shaded region is excluded by the di-jet resonance searches.
As discussed above, the signal rate depends almost exclusively on $m_{\tilde
  \tau_1}$ unless $|\lambda_{i11}^\prime| \ll 1$. 
As can be seen, in order to explain the di-photon excess
the lightest stau mass must be within the narrow window $375\,{\rm GeV} \leq
m_{\tilde \tau_1} \lsim 389\,{\rm GeV}$. 
The lower mass limit is required to forbid the two-body decay mode,
$\tilde \nu_i \to \tilde \tau_1^+ \tilde \tau_1^-$. Above this kinematical
threshold, one observes gradual suppression with stau mass due to gradual
decoupling. 
For a smaller value
  of $A_{i33}$ or $\theta$, the upper limit on the stau mass becomes stronger
  and the green-shaded region in Fig.~\ref{fig:RPV0} shrinks, until we have no
  allowed region left for $A_{i33}<10 \: m_{\tilde\tau_1}$ or for
  $\theta<\pi/7$.  
Contrary to the di-photon rate, the di-jet constraint is sensitive to
$|\lambda_{i11}^\prime|$ 
and excludes the region where $|\lambda_{i11}^\prime| > 0.08$.  

The dashed contours in Fig.~\ref{fig:RPV0} show the total decay width of the sneutrino in GeV.
As we discussed previously, the total width is dominated by the $\tilde \nu_i \to d \bar d$ mode and
depends only on $|\lambda_{i11}^\prime|$ unless $|\lambda_{i11}^\prime| \ll 1$.
As can be seen, $\Gamma_{\rm tot} > 300$ MeV is excluded by the di-jet
constraint in the region favoured by the di-photon excess. 
This is a prediction of the model in its minimal version: {\it if the signal persists and the resonance is
better resolved, it should have a narrow width.}

\section{Low-energy Constraints} \label{sec:low}
We must make sure that the di-photon favoured range of $\lambda'_{i11}$ in Fig.~\ref{fig:RPV0} is consistent with other low-energy constraints, such as electroweak precision observables and lepton flavour violating processes~\cite{Barbier:2004ez}.\footnote{For detailed discussions of the indirect constraints on the scalar di-photon resonance, see Refs.~\cite{Goertz:2015nkp, Staub:2016dxq}.} For instance, the constraint from charged current universality in lepton and quark sectors implies~\cite{Herz:2002gq, Dreiner:2006gu} 
\begin{align}
\lambda'_{11k} \ \leq \ 0.02\left(\frac{m_{\tilde d_{kR}}}{100~{\rm GeV}}\right).
\end{align}
Similar limits on $\lambda'_{11k}$ are also obtained from atomic parity violation in$^{133}$Cs~\cite{Barbier:2004ez}. 
From neutrino-lepton elastic scattering mediated by neutral currents, we get~\cite{Barbier:2004ez} 
\begin{align}
\lambda'_{21k} \ \leq \ 0.15\left(\frac{m_{\tilde d_{kR}}}{100~{\rm GeV}}\right), \qquad \lambda'_{2j1} \ \leq \ 0.18\left(\frac{m_{\tilde d_{jL}}}{100~{\rm GeV}}\right).
\end{align}

Large $\lambda'_{i11}$ interactions can also induce sizable lepton flavour violating radiative decays of charged leptons~\cite{deGouvea:2000cf}. Using the most stringent constraint from MEG on ${\rm BR}(\mu\to e\gamma)<5.7\times 10^{-13}$ at 90\% C.L.~\cite{Adam:2013mnn}, we obtain
\begin{align}
|\lambda'_{2jk}\lambda'^*_{1jk}| \ \lesssim \ 1.6\times 10^{-5} \left(\frac{m_{\tilde d_{kR}}}{100~{\rm GeV}}\right)^2.
\end{align}

Limits on $|\lambda'_{i11}|$ were also set from the electric dipole
  moment (EDM) constraints~\cite{Frank:1997aj}. Using the current best upper
  limit on electron EDM, $|d_e|<8.7\times 10^{-29}e.{\rm cm}$ at 90\%
  C.L.~\cite{Baron:2013eja}, we get $|\lambda'_{111}|\leq 9.3\times 10^{-6}$,
  whereas from muon EDM, we get a much weaker constraint:
  $|\lambda'_{211}|\leq 0.5$, assuming all the relevant squark and slepton
  masses in the loop to be 100 GeV. These limits can however be completely evaded by a suitable choice of phases in the squark mixing matrix or at least siginificantly
  weakened by making the squarks heavier: already if they are placed at 1
  TeV, the above constraints leave plenty of room for the values of
  $\lambda'_{i11}<0.1$ that we require to explain the di-photon resonance. 

For $i=1$, the $\lambda'_{111}$ coupling is also constrained by neutrinoless double beta decay ($0\nu\beta\beta$)~\cite{Mohapatra:1986su, Hirsch:1995ek}. Using the current 90\% CL combined limit on the $0\nu\beta\beta$ half-life for $^{76}$Ge isotope from GERDA phase-I: $T_{1/2}^{0\nu} > 3.0\times 10^{25}$ yr~\cite{Agostini:2013mzu}, we find~\cite{Faessler:1996ph, Allanach:2009xx}
\begin{align}
|\lambda'_{111}| \ \lesssim \  4.5 \times 10^{-4}\left(\frac{m_{\tilde{e}_L}}{100~{\rm GeV}}\right)^2\left(\frac{m_{\tilde{\chi}_1^0}}{100~{\rm GeV}}\right)^{1/2} \ \simeq \  0.025\left(\frac{m_{\tilde{\chi}_1^0}}{100~{\rm GeV}}\right)^{1/2}
\label{eq:ndbd}
\end{align}
for a selectron mass of 750 GeV.  Comparing this with the di-photon favoured region in Fig.~\ref{fig:RPV0}, we find that the $0\nu\beta\beta$ constraint still allows some parameter space for the $i=1$ case as long as the the lightest neutralino is heavier than about 50 GeV. 
We also note that our scenario satisfies the constraints from $S$ and $T$ parameters measured at LEP, since the sleptons are heavier than $375$ GeV. For details, see Fig. 3 of Ref.~\cite{Cho:1999km}.

To summarize, the low-energy constraints depend on additional sparticle
masses not involved in the di-photon explanation, and are easily
satisfied by making the sparticles appropriately heavy enough, without
affecting the di-photon signal.

\section{Model tweaks \label{sec:tweaks}}

One way to increase $\Gamma_{\rm tot}$ would be to allow the sneutrino to have
other decay modes, $X$. 
If $\Gamma_X$ were as large as or larger than $\Gamma_{d \bar d}$, the cross
sections would scale as 
\beqn
\sigma(pp \to \tilde \nu_i \to \gamma \gamma) & \ \appropto \ & |\lambda^\prime_{i11}|^2 \Big( 
\frac{ \Gamma_{\gamma \gamma} }{ c |\lambda^\prime_{i11}|^2 + \Gamma_X } \Big) \,, \nonumber \\
\sigma(pp \to \tilde \nu_i \to d \bar d) & \ \appropto \ & |\lambda^\prime_{i11}|^2 \Big( 
\frac{ c |\lambda^\prime_{i11}|^2 + \delta_{1 n} \Gamma_X }{ c |\lambda^\prime_{i11}|^2 + \Gamma_X } \Big) \,, \nonumber \\
\sigma(pp \to \tilde \nu_i \to  X) & \ \appropto \ & |\lambda^\prime_{i11}|^2 \Big( 
\frac{ \Gamma_X }{ c |\lambda^\prime_{i11}|^2 + \Gamma_X } \Big)  \,,
\eeqn
where $c$ is some constant.
A few remarks can be made.
First of all, all processes depend on $\Gamma_X$.
Second, the di-photon rate now also depends on $\lambda_{i11}^\prime$.
Therefore, for $\Gamma_X > 0$, to compensate for the suppression in the
di-photon rate, a larger value of $|\lambda_{i11}^\prime|$ 
will be preferred. 
The suffix $n$ of the Kronecker delta is 0 except for $X = d_k \bar d_l$ with $d_k$ being one of $d, s, b$ 
(and $\bar d_l$ being one one of $\bar d, \bar s, \bar b$) 
excluding $X = d \bar d$.
These decay modes can be opened up by introducing a non-zero
$\lambda_{ikl}^\prime$ 
coupling for the $L_i Q_k \bar D_l$ operator in the superpotential in Eq.~\eqref{eq:wrpv}.
For $n=1$, the di-jet cross section is independent of $\Gamma_X$, whereas it is suppressed with $\Gamma_X > 0$ for the $n=0$ case.

Appropriate constraints on $\sigma(pp \to \tilde \nu_i \to X)$ should be taken
into account\footnote{For a comprehensive list of LHC probes on hidden sector, see
  e.g. Ref.~\cite{Jaeckel:2012yz}.}.
For example, if some of the neutralinos and charginos are lighter than $\tilde \nu_i$, 
one can consider $\tilde \nu_i \to \nu \tilde \chi_j^0$ and $\tilde \nu_i \to \ell^\pm \tilde \chi_j^\mp$.
In RPV scenarios, the $\chi_j^0$ and $\chi_j^\pm$ subsequently decay into jets and leptons
via RPV interactions and these processes may be observed as  multi-jet and/or
multi-lepton with or without large missing transverse momentum final states. 
Constraints on these processes depend on the details of the final state
particles and the masses of 
$\chi_j^0$ and $\chi_j^\pm$, but are typically more stringent than the di-jet
constraint. 
Another possibility is $X = b \bar b$ or $b \bar s \: (s \bar b)$.
The upper bound on the $b \bar b$ signal cross-section is about 1 pb from the
di-bottom resonance search \cite{Khachatryan:2015tra},  
whilst the latter does not have any other constraint apart from the
 di-jet constraint previously covered. 

Additionally, one could tweak the model 
to explain a wider peak by having multiple
sneutrino resonances, e.g.~$\tilde \nu_e$ and $\tilde \nu_\mu$, with slightly
different masses, $\Delta m \sim {\cal O}(10)$ GeV: at present, statistics are such that
one cannot resolve these two different masses with the ATLAS data presented,
however this tweak predicts that in the future, the double-peak structure would
be resolved (the di-photon invariant mass resolution is around 1$\%$ i.e.\ $\sim
7$ GeV).  

A comparison by ATLAS with the 8 TeV di-photon data and their
  interpretation in 
terms of a 750 GeV resonance implies that production by $d \bar d$ is
disfavored at the 2.1$\sigma$ level~\cite{Moriond}. If this tendency in the
data persists, we should include the contributions from strange or bottom quarks, either of which is more compatible with
the 8 TeV inferred rate. Thus, instead of assuming a non-zero $\lambda_{i11}'$ only, we
would also be considering non-zero $\lambda'_{ijk}$ where $j$ and/or $k$ are
greater than 1. Strange or bottom quarks, being non-valence, have lower parton
distribution functions to produce a 750 GeV sneutrino than down quarks and so
an increase in the value of $\lambda'_{ijk}$ as compared to $\lambda'_{i11}$
would be required in order to fit the data. 


\section{Conclusion and discussion} \label{sec:concl}

One can explain the di-photon excess via resonant sneutrino production whilst
remaining on the allowed side of other collider constraints. The model
contains a stau of mass anywhere from 375 to 389 GeV and a 750 GeV sneutrino. 

It is interesting to note that resonant left-handed slepton production has
been used to simultaneously explain the ATLAS di-boson excess at 2 TeV in LHC
Run I and the anomalous magnetic moment of the muon~\cite{Allanach:2015blv}.
It remains to be seen whether the $R-$parity violating MSSM has enough freedom
to simultaneously fit these measurements (which are also discrepant with 
SM predictions) {\em and}\/ the di-photon excess addressed here. We leave 
the investigation of this issue
to a future paper. 

The most pressing concern resulting from this and other works is: will the 750
GeV $\gamma \gamma$ excess persist in future Run II data? If the answer is
`yes', there are some ways to discriminate our proposal from the other many
new physics scenarios that have explained the excess. Firstly,
the
largest possible width we can get in this scenario is 0.3 GeV and so our
base-line model predicts
that the mild preference in the ATLAS data for a width of 45 GeV will not
persist. With larger statistics, we predict that the angular distributions for
the $\gamma \gamma$ final state should agree more with a spin zero resonance
produced by $q \bar q$ initial state (as opposed to $gg$). Unfortunately, the
$\gamma Z$ and $ZZ$ signal rates are probably too small to be seen at
the LHC, given that they are all suppressed by a factor of $10^4$ or
more compared to the $\gamma \gamma$ signal. However, the signal rates for
di-jets or $WW$ are non-negligible and while 
backgrounds are large, these channels remain a hope to verify the model. 
Charged slepton signals producing $W \gamma$ and $W Z$ are an additional
prediction of the model, at a mass very close to 750 GeV. 

\section*{Note Added}
In the final stages of preparation of this manuscript,
Ref.~\cite{Ding:2015rxx} appeared, presenting an 
explanation for the di-photon excess using the sneutrino and $R-$parity
violating supersymmetry, finding that light staus and smuons in the range
375-480 GeV are favored, and we note some overlap with our paper. However, they have not included the $WW$ decay mode of the sneutrinos and have assumed a much larger soft SUSY-breaking term $A_{i33}=10$ TeV, which is potentially dangerous for vacuum stability. 

\section*{Acknowledgments}
We thank Florian Staub and Martin Winkler for
helpful communications regarding vacuum stability. The work of B.C.A. has been
partially supported by STFC grant 
ST/L000385/1. The work of 
P.S.B.D. is supported in part by a TUM University Foundation Fellowship, as
well as by the DFG with grant RO 2516/5-1. SR acknowledges the support of the
Gordon and Betty Moore foundation during a Graduate Fellowship at the Kavli
Institute for Theoretical Physics where part of this work was
undertaken. This research was supported in part by the National Science
Foundation under Grant No. NSF PHY11-25915.

\bibliographystyle{apsrev}
\bibliography{draft}

\end{document}